# NMR Study of Magnetic Properties of the Staircase Kagomé Antiferromagnet PbCu$_3$TeO$_7$*

**DAI Jia(代佳)[1], WANG Peng-Shuai(王朋帅)[1], SUN Shan-Shan(孙珊珊)[1], PANG Fei(庞菲)[1], ZHANG Jin-Shan(张金珊)[2], DONG Xiao-Li(董晓莉)[3], YUE Gen(乐艮)[3], JIN Kui(金奎)[3], CONG Jun-Zhuang(丛君状)[3], Sun Yang(孙阳)[3], YU Wei-Qiang(于伟强)[1]****

[1] *Department of Physics, Renmin University of China, Beijing 100872, China*

[2] *School of Energy, Power and Mechanical Engineering, North China Electric Power University, Beijing 102206, China*

[3] *Beijing National Laboratory for Condensed Matter Physics, Institute of Physics, Chinese Academy of Sciences, Beijing 100190, China*

* YW is supported by the NSFC under Grant Nos 11374364, 11222433 and the National Basic Research Program of China (Grant Nos 2011CBA00112). JZ is supported by the NSFC and the Scientific Research Foundation for the Returned Overseas Chinese Scholars, State Education Ministry.

** Email: wqyu_phy@ruc.edu.cn

(Received  )

We report the first nuclear magnetic resonance (NMR) study on single crystals of staircase Kagomé antiferromagnet PbCu$_3$TeO$_7$ ($T_{N1}$~36 K). A Curie constant Θ~-140 K is obtained by a Curie-Weiss fit to the high-temperature Knight shift of $^{125}$Te. The hyperfine coupling constant is estimated as $^{125}A_{hf}$= -67 kOe/$\mu_B$, and a strong interlayer coupling among staircase Kagomé planes is suggested with such a large hyperfine coupling, according to the lattice structure. The $^{63,65}$Cu NMR spectra are found by the zero-field NMR at $T=$ 2 K, and the internal hyperfine field are estimated to be 10.3 Tesla and 9.6 Tesla, for the Cu(1) and the Cu(2) sites respectively in the lattice. A second type of zero-field NMR signal with a large RF enhancement is also seen after field-cycling through a high magnetic field, which probably characterizes quenched disorder in frustrated magnet.

The ground states and excitations of geometrically frustrated quantum magnets have attracted enormous interests in condensed matter physics. Particularly, two-dimensional (2D) S= 1/2 Kagomé Heisenberg antiferromagnet (KHAF) may have novel quantum disordered states, such as spin liquids, due to strong quantum fluctuation and magnetic frustration[1-4]. For systems with imperfect Kagomé structure, magnetic anisotropy induced by Dzyaloshinsky-Moriya (DM) interaction[5], spatially anisotropic exchange[6], and/or interlayer coupling may reduce geometry frustration and lead to antiferromagnetic ordering with very low Néel temperatures.

On the other hand, the interplay of geometric frustration, magnetic anisotropy, and quantum fluctuations produces competing ground states with complex magnetic structures, which may lead to potential multifunctional materials. Staircase Kagome lattice with buckled Kagomé layers is one example, which has been realized in several materials, including $A_3V_2O_8$ (A= Cu,Co,Ni)[7-9] and $PbCu_3TeO_7$[10]. Consecutive magnetic transitions and magnetism induced ferroelectricity were reported in $Ni_3V_2O_8$, characterizing a multiferroic material with strong magnetoelectric coupling[11,12]. The lattice structure of the newly discovered $PbCu_3TeO_7$ is illustrated in Fig.1. The planes are stacked along the crystalline *a* direction with Pb and Te atoms as spacer layers. The buckled Kagomé layer has two nonequivalent copper sites, Cu(1) in $CuO_6$ octahedrons and Cu(2) in $CuO_4$ tetrahedrons with a ratio of 2:1, each aligned in chains along the *b* direction. The magnetic couplings among neighboring Cu(1) spins and neighboring Cu(1) and Cu(2) spins are not uniform through corner and edge sharing oxygen[10], and therefore the system deviates from an ideal Kagomé lattice antiferromagnet. It has been reported that the Weiss constant $\Theta$ obtained from the magnetic susceptibility is about -140 K, whereas consecutive antiferromagnetic transitions occur at much lower temperatures with $T_{N1}$~36 K, $T_{N2}$~25 K, and $T_{N3}$~17 K, and their respective magnetic structures are still not resolved yet [10].

In this paper, we present the first NMR study on $PbCu_3TeO_7$, which is helpful for understanding its magnetic properties microscopically. Our high-temperature $^{125}$Te Knight shift data in the paramagnetic phase reveals a large hyperfine coupling on the $^{125}$Te sites. This suggests a strong interlayer hyperfine coupling, which should help the magnetic ordering of the system. Below the Neel temperature, the zero-field NMR spectra are consistent with Cu signals from both the Cu(1) and the Cu(2) sites. In addition, we also found a second zero-field NMR signal, which has a large RF enhancement and is only detectable after field cycling through high magnetic fields. Here the field cycling refers to a process that a finite magnetic field ($H_{cycle}$) is first switched on and then reduced to zero at the same temperature. We discuss that the second signal is likely from quenched disorder which is ferromagnetic in frustrated magnet.

The PbCu$_3$TeO$_7$ single crystals were grown by the flux growth method with NaCl/KCl as flux[10]. Our X-ray diffraction (data not shown) and susceptibility data (data shown later) are consistent with the literature[10]. The high-temperature bulk susceptibility was measured in a superconducting quantum interference device (SQUID) magnetometer, and the low-temperature magnetization was measured in a vibrating sample magnetometer (VSM). NMR measurements were conducted on single crystals with mass ~ 3 mg, with field applied along the crystalline [0 1 1] direction at the paramagnetic phase and under zero field in the ordered phase. The NMR spectra were collected by the standard spin-echo sequence. The high-temperature $^{125}$Te spectra were obtained by fast Fourier transform (FFT) of the spin-echo signal. The Knight shift in the paramagnetic phase was calculated with $^{125}K = f/^{125}\gamma H - 1$, where $f$ is the center frequency of the resonance line, $H$ is the external field, and $^{125}\gamma$ = 13.454 MHz/T is the gyromagnetic ratio. The low-temperature broad zero-field $^{63,65}$Cu spectra were obtained by integrating the spin-echo intensity with frequency swept through the resonance line.

The $^{125}$Te ($I$ = 1/2) NMR spectra, with temperatures from 275 K down to 37 K, are first shown in Fig.2 (a). Upon cooling, the spectra shift to lower frequencies and their full width at half maximum (FWHM) increases significantly (~40 KHz at 275 K and ~ 300 KHz at 37 K). The NMR signal is too broad to be detectable with temperature down to 36 K, which is an indication of magnetic transition, consistent with the $T_{N1}$ detected by the susceptibility measurements.

The Knight shift $^{125}K$ is calculated and shown as a function of temperature in Fig. 2(b). $^{125}K$ is close to zero at 275 K and decreases with temperature as cooling, indicating a negative hyperfine coupling transferred from copper ions[13]. The Knight shift above 100 K is well fit with a Curie-Weiss form, $^{125}K(T)$=A+C/($T$-Θ), as shown by the dashed line in Fig. 2(b). The fitting gives Θ~-140 ± 20 K, which is much higher than the Néel temperature, and therefore strong magnetic frustration is suggested.

In Fig. 2 (b), we further plot our bulk susceptibility data χ of the crystal, measured as a function of temperature with a 1 T field. The χ also shows a Curie-Weiss increase upon cooling, consistent with the reported data [10]. In the inset of Fig. 2(b), $^{125}K$ is plotted against χ with temperature as an implicit parameter, where a linear relation between them is clearly seen. From the formula $^{125}K=\dfrac{^{125}A_{hf}}{N_A\mu_B}\chi$, where the $\mu_B$ is Bore magneton and $N_A$ is the Avogadro's constant, the hyperfine coupling constant is calculated to be $^{125}A_{hf}$= -67 kOe/$\mu_B$. Such a large hyperfine coupling indicates that Te is strongly coupled to Cu moments. According to previous calculations of the electronic structure[10], the Te atoms do not participate in any intralayer hopping path; however, they mediate all the paths perpendicular to the Kagomé planes. Among all interlayer hoping paths, the strongest one is Cu(1)-O-Te-O-Cu(2), as shown by the dashed lines in the inset of Fig. 2(a), where two Cu atoms are from neighboring Kagomé planes. Given such a large hyperfine coupling observed on the $^{125}$Te sites, our data suggests that interlayer coupling is strong in this material. As a result, the strong interlayer magnetic coupling should help the three-dimensional magnetic ordering with a high Néel temperature (36 K).

We did not find the Cu NMR signal spectra in the paramagnetic state, possibly because of the strong magnetic fluctuations on the Cu sites. However, at temperatures far below the Néel transition, a zero-field (ZF) NMR spectrum is obtained from 65 MHz to 145 MHz at $T$= 2 K, after zero-field cooling (ZFC). The spectrum is marked as $A$1 to distinguish with the second NMR signal we describe later. The observation of the ZF spectrum indicates the large internal hyperfine fields due to magnetic ordering. The $A$1 spectra have been checked by NMR measurements on several crystals, and their lineshapes are consistent.

The NMR spectrum is very broad, with a finite background and six peak-like features from 65 MHz to 140 MHz. With such a large resonance frequency and the multiple peak feature, we think that the signal is unlikely from $^{125}$Te which is a spin-1/2 nuclei. On the other hand, considering the Cu(1) and the Cu(2) sites in this staircase Kagomé lattice, and two types of spin-3/2 copper nuclear isotopes ($^{63}$Cu and $^{65}$Cu), twelve Cu NMR lines are expected with four center transitions and eight satellite transitions. Therefore, it is reasonable to attribute the observed NMR spectrum from Cu isotopes. In particular, we assigned the narrows lines to the center transitions which have weak second order quadrupolar corrections, comparing with the satellite lines with the first order corrections, due to lattice inhomogeneity. Given the gyromagnetic ratio of $^{63}\gamma$=11.285 MHz/T and $^{65}\gamma$=12.099 MHz/T, and the natural abundance of 69% and 31% for $^{63}$Cu and $^{65}$Cu respectively, we first tentatively associate the observed narrow NMR peaks to four center transition as shown by the arrows labeled with $^{63}$Cu(1c), $^{65}$Cu(1c), $^{63}$Cu(2c), and $^{65}$Cu(2c) respectively. This is roughly consistent with the theoretical value of the relative spectral weight $^{63}$Cu(1c): $^{65}$Cu(1c): $^{63}$Cu(2c): $^{65}$Cu(2c) ≈ 128 : 62 : 69 : 31, and two hyperfine fields on the Cu(1) and the Cu(2) sites respectively. However, we have difficulty to resolve all satellite lines. Nevertheless we attempted to assign several peaks to the satellite transition as labeled in Fig.3, given the quadrupole moments of 0.211 and 0.195 barns for $^{63}$Cu and $^{65}$Cu respectively.

The internal hyperfine field on Cu(1) and Cu(2) sites are estimated to be $B_{in}(1)$ ~ 10.3 Tesla and $B_{in}(2)$ ~ 9.6 Tesla respectively. Here we neglected the second-order quadrupolar correction to the center lines, which is in an order of 2 MHz estimated from the line splitting of the satellites. We note that $A$1 spectrum does not show hysteresis with temperature or field changes, and no obvious RF enhancement is observed by comparing the NMR in the paramagnetic phase, as expected for bulk NMR signal in the ordered antiferromagnetic phase.

In the following, we show a second ZF NMR signal with a large RF enhancement, which is created at low temperatures after field cycling through a high magnetic field. As shown in Fig. 4(a), the zero-field spectrum at $T$ =2 K, labeled as $A$2, is obtained after field cycling under a 6 Tesla field and the optimized RF excitation power is 22 dB lower than that of the $A$1 spectra under the same pulse length. Therefore, a RF enhancement factor η~20 is estimated for the $A$2 signal. Here the same vertical scale is used for Fig. 3 and Fig.4(a) for comparison.

The $A2$ signal is not seen after zero-field cooling from high temperatures, whereas $A1$ does not change under different field or thermal treatments. The hysteresis of the $A2$ spectrum is further measured under a magnetic field $H$ and with different cycling fields $H_{cycle}$. As shown in Fig. 4(b), after field cycling under 6 T field, the measurement is taken under external magnetic field $H$ from 0 to 0.4 T, the echo intensity at 116 MHz (center of the spectrum) is quickly suppressed to zero. In Fig. 4(c), the echo intensity is shown as a function of cycling field $H_{cycle}$, while the external magnetic field remains zero. The echo intensity is zero with $H_{cycle}$ from 0 to 2 T, then starts to increase rapidly with $H_{cycle}$ from 3 T to 6 T, and saturates with $H_{cycle}$ above 6 T. Clearly, the signal strength is enhanced after field cycling through a threshold field (6 T), but suppressed under a small external magnetic field.

Comparing with $A1$, $A2$ is broadly distributed above 40 MHz and is peaked at ~115 MHz, and the signal intensity for $A1$ and $A2$ is about the same magnitude. The spin-spin relaxation time $T_2$ is about 14 $\mu$s for $A2$ at 2 K, much shorter than that of $A1$ with $T_2$ ~60 $\mu$s (data not shown). Therefore, strong magnetic fluctuation is suggested for the $A2$ spectrum. Therefore, $A2$ is characterized by nuclei environments with strong thermal and field hysteresis, a large RF enhancement, strong magnetic fluctuations, and inhomogeneous hyperfine fields. Since NMR is a local probe, the differences in the field hysteresis and different $T_2$ between $A1$ and $A2$ suggest different regions of the sample.

It is known that zero-field NMR in ordered ferromagnet has large RF enhancement, fast relaxation, and broad linewidth due to domain wall dynamics, and the signal intensity is suppressed under field[14]. Our $A2$ signal is consistent with the NMR signal from ferromagnet domains, although bulk $PbCu_3TeO_7$ is an antiferromagnet. In fact, we show that the $A2$ signal does not represent bulk properties of the sample. First, if we normalize the NMR intensity by the RF enhancement factor, the intensity of the $A2$ signal at 116 MHz is about 5% of that of $A1$. Second, the magnetization data measured at 2.2 K, as shown in Fig. 5, increases linearly with field from 0 T to 12 T. Therefore, no evidence for field induced spin flop or other magnetic transition in the bulk sample to account for the field cycling induced NMR signal.

In fact, our $A2$ signal is consistent with quenched disorder in a frustrated antiferromagnet. In the frustrated system, the ground states are highly degenerate [17-19], and net ferromagnetic moment may be created close to quenched disorder. In this case, an external field above a threshold field polarizes the defects and/or the neighboring spins, and therefore creates small ferromagnetic domains surrounding the defects in the antiferromagnetic background. When the cycling field is removed, the remnant ferromagnetism leads to a large zero-field NMR signal. Since the domain size is small, the NMR signal from this region should have a short $T_2$ and a large RF enhancement as seen in regular ferromagnet. We indeed observed similar field cycling induced NMR signal in other frustrated magnets, including $CuBr_2$ and $FeVO_4$ (data not shown). Therefore, the field cycling induced NMR signal is likely a characteristic feature for frustrated magnet with quenched disorder.

We note that field cycling induced NMR signal was reported in spin glass system Cu-Mn alloy[15] and frustrated antiferromagnet $TbMn_2O_5$[16]. In the case of Cu-Mn alloy, field cycling induces remnant magnetization and results in enhanced bulk NMR signal from coherence of neighboring spins[15]. For TbMn$_2$O$_5$, the enhanced NMR signal is attributed to field cycling induced ferromagnetic ordering of Tb moments[16]. These observations, including in our case, are consistent that remnant ferromagnetism plays an essential role for the induced NMR signal.

In summary, we studied the magnetic properties of frustrated staircase Kagomé compound PbCu$_3$TeO$_7$ by NMR. Strong interlayer coupling is suggested by the large $^{125}$Te Knight shift in the paramagnetic phase. We also identified the hyperfine fields on the Cu(1) and the Cu(2) sites, which gives microscopic information for the magnetic structure. A second type NMR signal is also seen by a field cycling process, which may be a characteristic feature for frustrated magnet with quenched disorder.

**Figures Caption**

**Fig. 1:** (color online) Crystal structure of PbCu$_3$TeO$_7$. Cu(1) in CuO$_6$ octahedrons and Cu(2) in CuO$_4$ tetrahedrons (oxygen not drawn) form the buckled Kagomé lattice layer. Te and Pb are spacers to separate the Kagomé layers.

**Fig. 2:** (color online) (a) High-temperature NMR spectra of $^{125}$Te with a fixed field *H*= 11.5 T, applied along the crystalline [0 1 1] direction. The vertical dashed line denotes the reference frequency at zero Knight shift. Inset: the major interlayer hopping path Cu(1)-O-Te-O-Cu(2) denoted by the dashed line. (b) The temperature dependence of the Knight shift $^{125}K$ (squares) and the bulk susceptibility χ (circles). The blue dashed line is the fit of the $^{125}K$ to the function $^{125}K$ (*T*)

=A+C/($T$-Θ) above 100 K. Inset: $^{125}K$ plotted against magnetic susceptibility χ above $T_{N1}$, where the straight line is a linear fit to the data.

**Fig. 3:** (color online) (a) The zero-field Cu spectrum measured at 2 K with a normal RF excitation power ($A$1, solid squares). The blue and green arrows denote the signal from the Cu(1) and the Cu(2) sites respectively. The subscript 'c' represents the central line and 's' represents the satellite line.

**Fig. 4:** (color online) (a) The zero-field spectra measured at 2 K with a low rf power and after a 6 T field cycling ($A$2, open circles). The vertical scale is the same as Fig.3. (b) The echo intensity of $A$2 after a 6 T field cycling, as a function of measurement field with fixed resonance frequency $f$ = 116 MHz. (c) The echo intensity of $A$2 as a function of the cycling field $H_{cycle}$ with fixed $f$ = 116 MHz and with zero measurement field.

**Fig. 5:** The magnetization curve measured at 2.2 K with field from 0 up to 12 T applied along the [0 1 1] crystalline direction.

**Fig. 1**

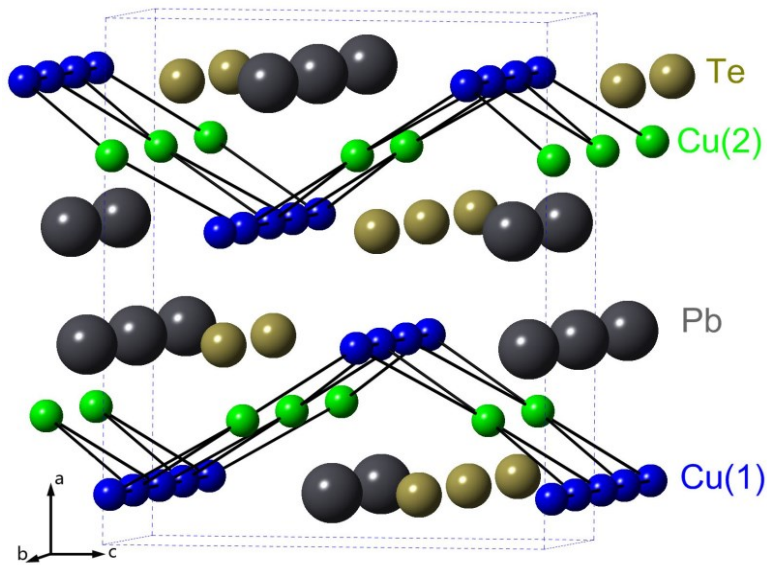

**Fig. 2**

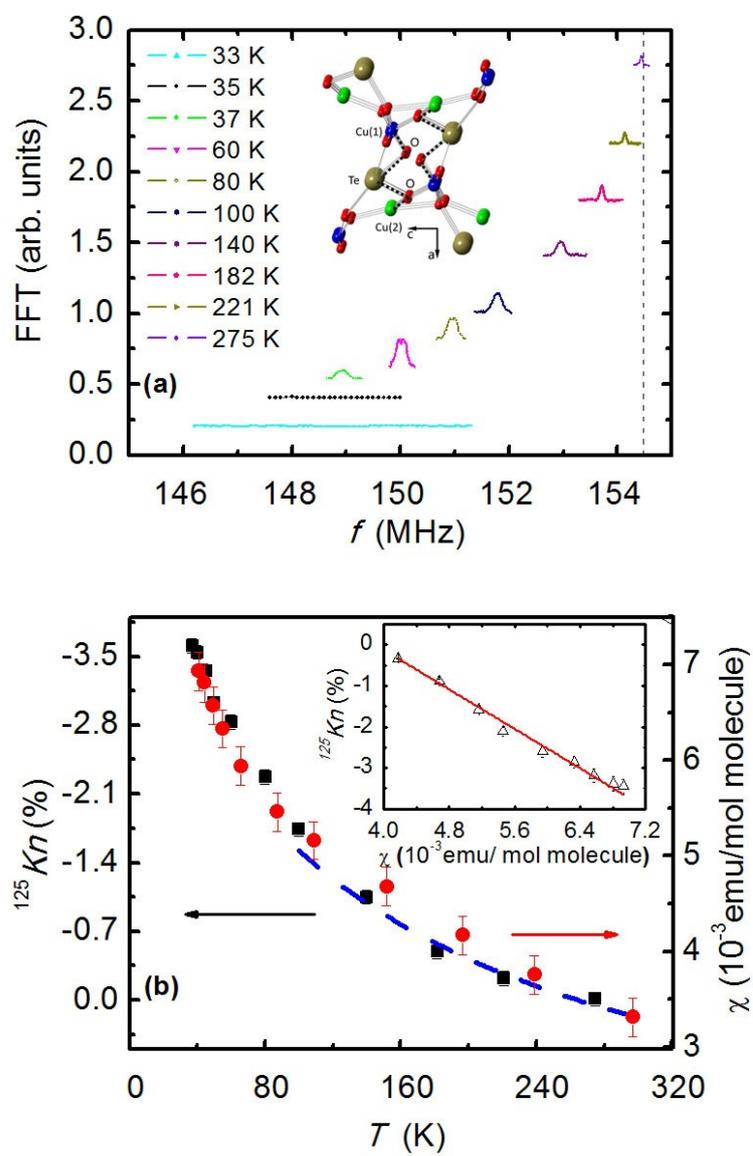

**Fig. 3**

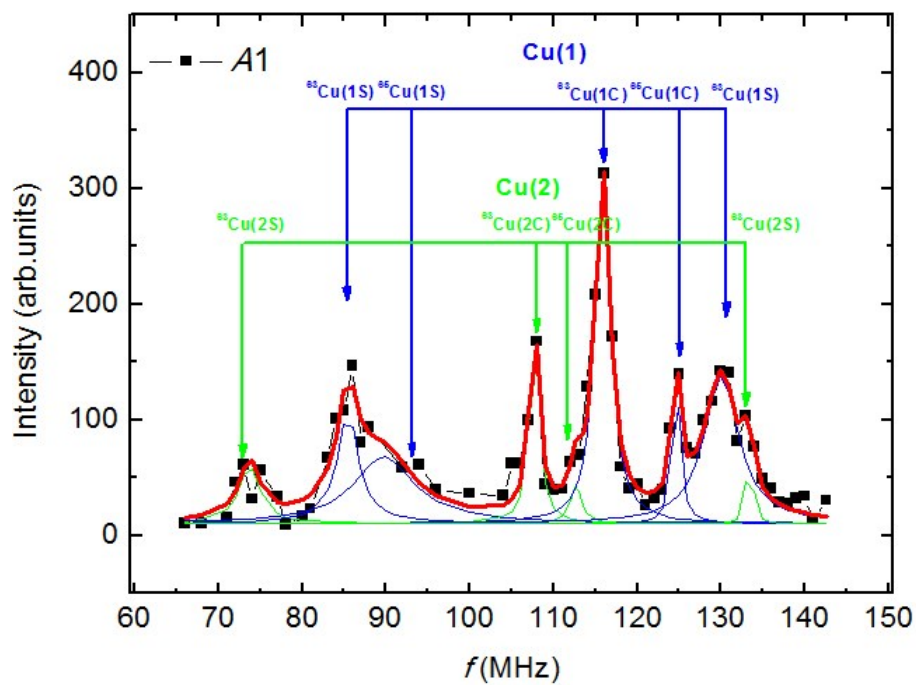

**Fig. 4**

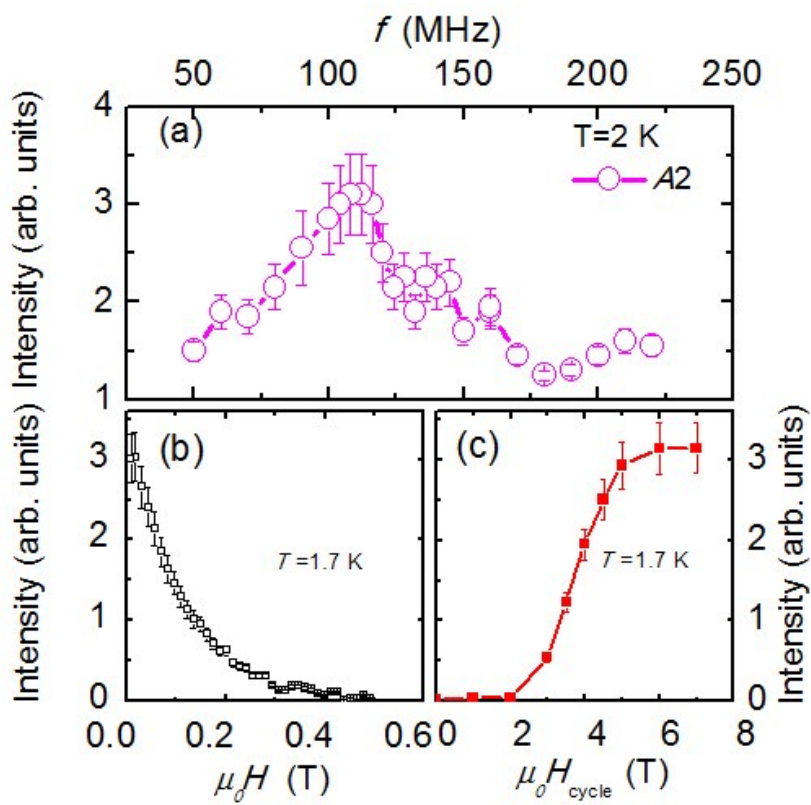

**Fig. 5**

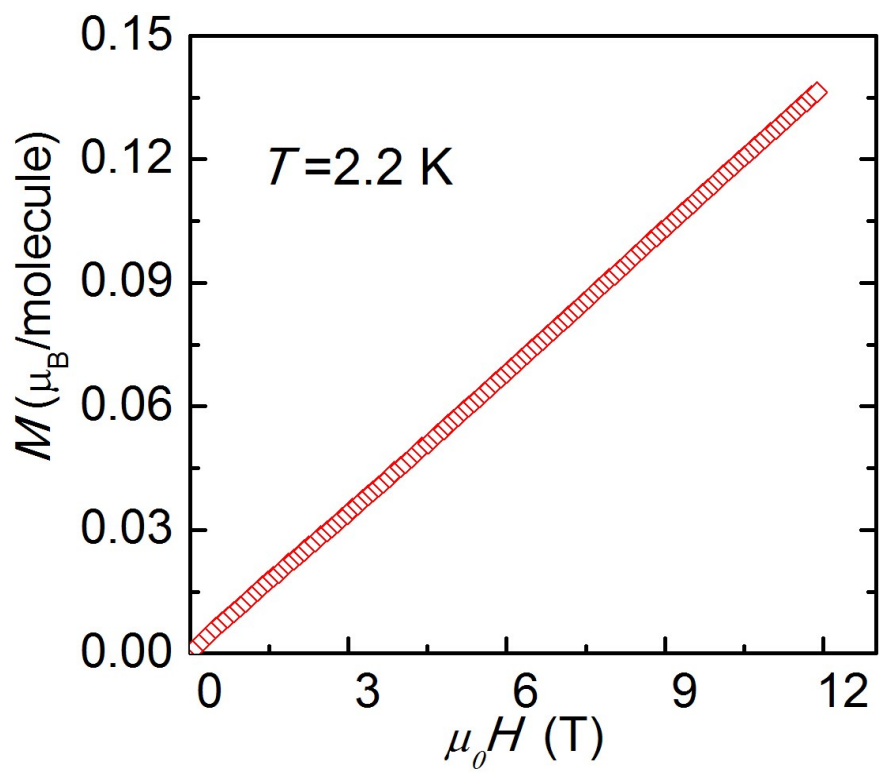